\documentclass[aps,prl,twocolumn,groupedaddress,showpacs]{revtex4}

\usepackage{graphicx}
\usepackage{color}
\definecolor{brown}{rgb}{.6,0,0}
\definecolor{grey}{rgb}{.5,.5,.5}
\definecolor{dgreen}{rgb}{0,.6,0}

\newcommand{\Rey}{\mathrm{Re}}
\newcommand{\eg}{e.g. }
\newcommand{\D}{\displaystyle}

\newcommand{\ri}{\mathrm i}
\newcommand{\rd}{\mathrm d}

\begin{document}


\title{Shark skin effect in creeping films}


\author{M. Scholle}
\author{N. Aksel}
\email[]{tms@uni-bayreuth.de}
\affiliation{Department of Applied Mechanics and Fluid Dynamics, University of Bayreuth,
 Universit\"atsstra\ss e 30, D--95440 Bayreuth, Germany}


\date{\today}

\begin{abstract}
If a body in a stream is provided with small ridges aligned
in the local flow direction, a remarkable drag reduction can
be reached under turbulent flow conditions. This surprising
phenomenon is called the 'shark skin effect'. We demonstrate, that
a reduction of resistance can also be reached in creeping flows
if the ridges are aligned perpendicular to the flow direction.
We especially consider in gravity--driven film flows the effect
of the bottom topography on the mean transport velocity.
\end{abstract}


\maketitle

\section{Introduction}
It is yet controversely discussed that the dermal surface morphology
of sharks is in order to improve the sharks' swimming performance~\cite{Vogel}.
Nevertheless, it is widely accepted that for bodies in turbulent flows a reduction of
skin friction by some percent can be reached if the surface of the body
is provided with small ridges aligned in the local flow
direction~\cite{Dinkelacker,Dinkelacker02}.
This rather counter--intuitional phenomenon occurs in turbulent flows if
the riblet spacing is smaller than the typical diameter of the streamwise
vortices such that the vortices are forced to stay above the riblets and
touch only the tips of the riblets.

The criterion for turbulent flow conditions is based on the Reynolds number
\begin{equation}                                        \label{Re}
 \Rey := \frac{\varrho UL}{\eta}\;,
\end{equation}
which gives a comparison between inertial forces
and friction forces. By $U$ and $L$ a characteristic
flow velocity and a characteristic length of the system are denoted,
$\varrho$ and $\eta$ are the mass density and the dynamic viscosity of the
fluid. At high Rey\-nolds numbers, typically $10^4$ or larger, the flow is
turbulent and vortices are created due to inertia. At small Rey\-nolds number
the flow is laminar and at vanishingly small Reynolds numbers $\Rey\ll 1$
the flow is creeping. Hence, in creeping flows a 'shark skin effect' is
not expected since inertia--induced vortices are absent.

\begin{figure}
\includegraphics[width=.45\textwidth]{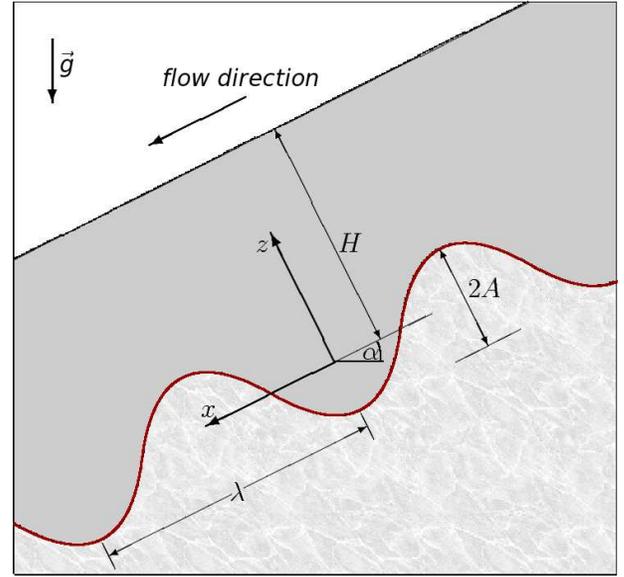}
\caption{\label{sketch}Sketch of the flow geometry}
\end{figure}

If the the riblets are aligned perpendicular to the flow direction,
however, kinematically induced vortices can be created even in creeping
flows where inertia is absent~\cite{Pozrikidis,waviness}.
We especially consider a steady, gravity--driven film flow
of an incompressible Newtonian fluid on an inclined plane. The bottom
of the inclined plane is provided
with periodic corrugations according to FIG.~\ref{sketch}.
The mean film thickness is denoted by~$H$, the periodic length
by~$\lambda$ and the mean inclination angle by~$\alpha$.
A Cartesian coordinate system is used with the $x$--axis placed at the mean
level of the bottom contour, the $y$--axis in line with the ridges
and the $z$--axis normal to the mean level of the bottom.

\section{Mathematical formulation}

\subsection{Basic equations and solving procedure}
The basic field equations are the continuity equation and
the Navier--Stokes equations~\cite{Spurk}
\begin{eqnarray}
 \nabla\cdot\vec v &=& 0 \;,                   \label{continuity}\\
 \varrho\left[\frac{\partial\vec v}{\partial t}
 +\left(\vec v\cdot\nabla\right)\vec v\right]
   &=& -\nabla p+\eta\nabla^2\vec v+\varrho\vec g\;. \label{NSt}
\end{eqnarray}
By~$\vec g$ the gravity acceleration is denoted.
The bottom contour is in general characterised by a $2\pi$--periodic
function~$b(\tilde x)$ as $z=b\left(2\pi x/\lambda\right)$.
Along the bottom the flow fulfills the no--slip condition 
\begin{equation}
 {\vec v} = \vec 0\;.                    \label{Ha_allgemein}
\end{equation}
Assuming that the film height exceeds the wavelength of the bottom,
the curvature of the free surface can be neglected~\cite{Wierschem,waviness}.
Then, the free surface is given as $z=H$. At the free surface the dynamic
boundary condition
\begin{equation}                                       \label{Dy_allgemein}
 \D\vec{e}_z  \left\{ \left[\tilde p - {\tilde p}_s
          \right] \underline{\underline 1}
         - \eta \left[\nabla \otimes \vec{v} +
         (\nabla \otimes \vec{v})^T\right] \right\} = \vec 0\;,
\end{equation}
which is the equilibrium between pressure jump at the surface
and viscous forces, has to be fulfilled.
By $p_s$ the pressure of the surrounding is denoted.
If the spatial extensions of the bottom in $y$--direction
are sufficiently large, side wall effects can be neglected
and a two--dimensional flow geometry can be assumed.
This allows for a representation of the velocity field
in terms of a stream function, i.e.
\begin{equation}
 \vec v = \frac{\partial\psi}{\partial z}{\vec e}_x
       -\frac{\partial\psi}{\partial x}{\vec e}_z \label{2D-Ansatz}
\end{equation}
by which the continuity equation~(\ref{continuity}) is identically fulfilled.
Since due to the creeping condition $\Rey\ll 1$ the inertia terms
at left hand of the Navier--Stokes equations are negligible,
EQ.~(\ref{NSt}) simplifies to the Stokes equations
\begin{equation}
 \vec 0 = -\nabla p+\eta\nabla^2\vec v+\varrho\vec g\;.
\end{equation}
Reconsidering EQ.~(\ref{2D-Ansatz}), the solution of Stokes
equations is explicitly given as~\cite{waviness}
\begin{eqnarray}
 \psi &\!=& \!\frac{\varrho gH^3\sin\alpha}{2\eta}\left[
    \frac{z^2\!}{H^2}-\frac{z^3\!}{3H^3}+\Re\!\left[r\!\left(\xi\right)
     +\frac{z}{H}q\!\left(\xi\right)\right]\right],\quad  \label{sol_complex}\\
 p &\!=& \! p_s  
      + \varrho g\left[\left(H-z\right)\cos\alpha+\frac{\pi H^2\sin\alpha}{
         \lambda}\Im q'\!\left(\xi\right)\right]\,, \label{p_complex}
\end{eqnarray}
where $q(\xi)$ and $r(\xi)$ are
holomorphic functions of the complex variable
\begin{equation}                              \label{xi:=}
    \xi := \pi\frac{z+\ri x}{\lambda}\;.
\end{equation}
The symbols $\Re$ and $\Im$ denote the real and imaginary part of
a complex expression, the prime derivation with respect to~$\xi$.
By inserting the general solution~(\ref{sol_complex}, \ref{p_complex})
in the boundary conditions~(\ref{Ha_allgemein}, \ref{Dy_allgemein}), a
set of equations for the boundary values of the two holomorphic
functions $q(\xi)$ and $r(\xi)$ is derived. Due to the periodicity of
the flow a representation of the boundary values of $q(\xi)$ and $r(\xi)$
by Fourier series can be applied, which leads to an algebraic set
of equations for the series coefficients. After truncating to a finite
number of Fourier modes, its solution is determined by using computer algebra,
\eg {\sc Maple}. The above
method and the solving procedure
\color{black}
is described in detail in~\cite{waviness}.
For the calculations presented here the series have been truncated
at the mode number where the values of the coefficients fall below~$10^{-10}$.
This leads to truncation orders between $12$ and $36$.
\color{black}

\subsection{Bottom shapes}
\begin{figure}
\includegraphics{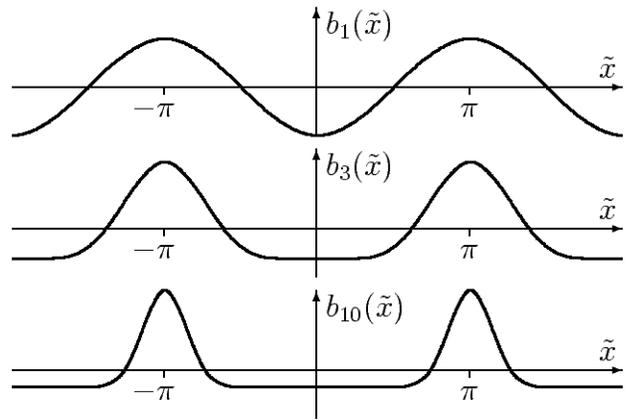}
\caption{\label{shapes}The three different bottom shapes $b_1$, $b_3$
 and $b_{10}$.}
\end{figure}

For the modelling of wall roughness Panton~\cite{Panton} suggests a
'brush model', i.e. an infinite array of equidistant narrow peaks.
We approach this strongly idealised shape by three different
trigonometric polynomials as shown in FIG.~\ref{shapes}.
At first, we consider a harmonic contour, i.e.
\begin{equation}                                     \label{DefPT1}
 b_1(\tilde x)=-A\cos\tilde x\;,
\end{equation}
with amplitude~$A$ as a comparatively smooth shape. In contrast to this,
by the third order trigonometric polynomial
\begin{equation}                                     \label{DefPT3}
 b_3(\tilde x) = -\frac{A}{16}\left[15\cos\tilde x-6\cos 2\tilde x
                  +\cos 3\tilde x\right]
\end{equation}
an array of peaks is given. Since the derivatives of $b_3(\tilde x)$
up to the $5$th order vanish, a flat region between the peaks becomes
apparent. The parameter~$A$ is again the amplitude, in the sense
that~$b_3(\pi)-b_3(0)=2A$.
The third shape shown in FIG.~\ref{shapes}, which is defined as
\begin{eqnarray}
 b_{10}(\tilde x) &=& -\frac{A}{262144}\Big[167960\cos\tilde x
                      -125970\cos 2\tilde x\nonumber\\[.5ex]
 &&\hspace*{-1.5cm} +77520\cos 3\tilde x\!-\!38760\cos 4\tilde x
                     \!+\!15504\cos 5\tilde x\!-\!4845\cos 6\tilde x\nonumber\\
 &&\hspace*{-1.5cm} +1140\cos 7\tilde x -190\cos 8\tilde x +20\cos 9\tilde x
                      -\cos 10\tilde x\Big]\,,   \hspace*{-1cm}\label{DefPT10}
\end{eqnarray}
is an array of significantly narrower peaks than in the case~$b_3(\tilde x)$.
Note, that the derivatives of~$b_{10}(\tilde x)$ vanish up to the $19$th
derivative. Therefore, $b_{10}(\tilde x)$ is a good approximation for
the idealised array of brush--like peaks by a continuous bottom shape.

\section{Results}

\subsection{Streamlines and vortex creation\color{black}}
\begin{figure}[floatfix]
\includegraphics{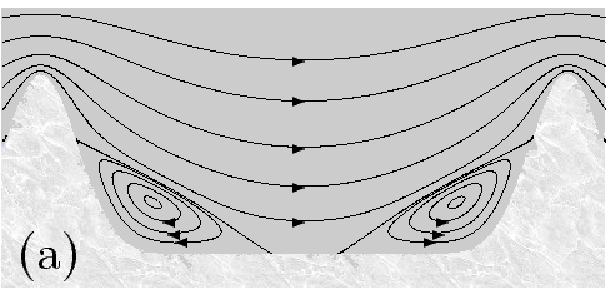}

\includegraphics{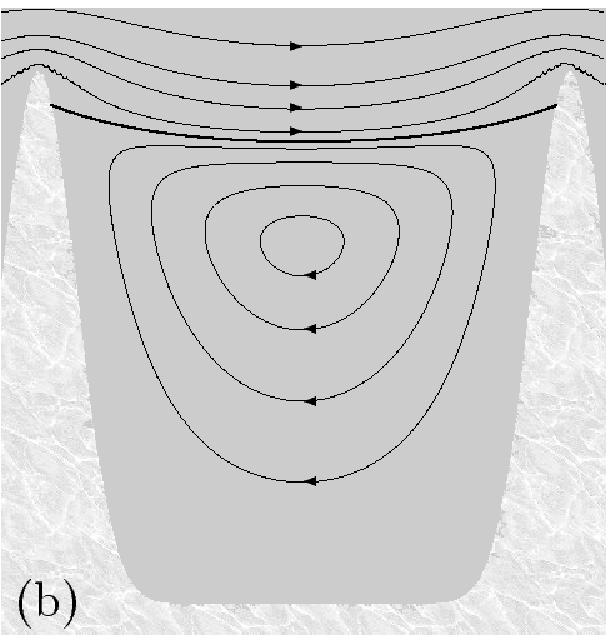}
\caption{\label{streamlines}Near--bottom streamlines of film flows
 over the shape~$b_{10}$ with amplitude (a) $A=0.17\lambda$
 and (b) $A=0.5\lambda$.}
\end{figure}

Streamline patterns have been calculated for various shapes, amplitudes
and film heights. For small amplitudes the streamlines follow the
bottom contour, whereas flow separation is observed if the amplitude
exceeds a critical limit.
As representative examples in FIG.~\ref{streamlines}
the streamlines in the vicinity of the bottom are presented for 
the shape~$b_{10}$ with two different amplitudes, namely
$A=0.17\lambda$ and $A=0.5\lambda$.
The film heights, $H=1.70\lambda$ in (a) and
$H=2.24\lambda$ in (b), are chosen in order to receive the same
flow rate in both cases.

The critical amplitude for the primary
flow separation is $A\approx 0.107\lambda$. Thus, 
\color{black}
in FIG.~\ref{streamlines}a flow separation is already apparent:
A vortex pair has been created at the positions of maximum curvature.
With increasing amplitude the vortices are growing, which leads
to the merging of the two vortices to a single one. Such a
case with a large single vortex which covers the major part
of the region between two neighbouring peaks is shown in
FIG.~\ref{streamlines}b. In this example we especially see
a slightly curved separatrix passing nearly from tip to tip.
This feeds the hope of a probable resistance reduction,
since the vortices are supposed to act like
'fluid roller bearings'.
By increasing the amplitude further, a secondary vortex pair is
created at the critical amplitude  $A\approx 0.606\lambda$ for
secondary flow separation.
\color{black}%

\subsection{Mean transport velocity}
We define the mean transport velocity as
\begin{equation}                \label{u_t:=}
 u_t:= \frac{\dot V}{H_t}\;,
\end{equation}
where the two--dimensional flow rate is given as
\begin{equation}
 \dot V = \int\limits_{b(2\pi x/\lambda)}^H u \rd z
        = \psi(x,H)-\psi\left(x,b(2\pi x/\lambda)\right)\;.
\end{equation}
The quantity~$H_t$, subsequently called
the mean transport thickness, has to be understood
as mean thickness of the part of the flow which contributes to
the material transport, i.e. the film
above the separation areas. If the separatrix of the
primary vortex is given as~$z=s(x)$, $x_1\leq x\leq x_2$,
the mean transport thickness results in
\begin{equation}                              \label{h':=}
 H_t:=H-\frac{1}{\lambda}\int\limits_{x_1}^{x_2}
 \left[s(x)-b(2\pi x/\lambda)\right] \rd x\;.
\end{equation}
\begin{figure}
\fbox{\includegraphics{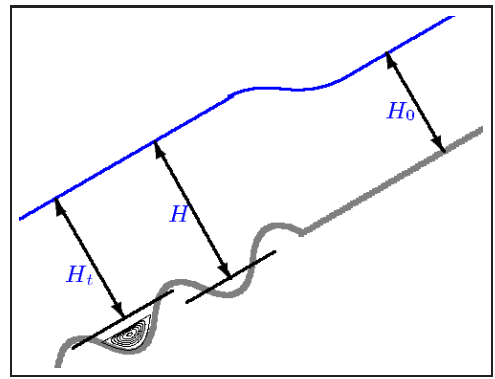}}
\caption{\label{hDefs}The quantities~$H_t$, $H$ and~$H_0$.}
\end{figure}
The quantity $H_t$ has to be carefully distinguished from the
mean geometrical film thickness~$H$ of the entire film 
which includes the vortices. Furthermore, by
\begin{equation}                              \label{h_0:=}
 H_0 := \left(\frac{3\eta\dot V}{\varrho g\sin\alpha}\right)^{\D\frac13}
\end{equation}
the reference thickness of a film flow on a plane bottom with the same flow
rate as the flow over the topography is defined~\cite{waviness}.
The three quantities $H_0$, $H$ and~$H_t$ are illustrated in FIG.~\ref{hDefs}.
Thus, for a fixed flow rate~$\dot V$, the comparison of the mean
transport thickness~$H_t$ with the reference thickness~$H_0$
delivers an adequate measure for enhancement or reduction of
the mean transport velocity in the film: In case of $H_t>H_0$, the mean
transport velocity is reduced, whereas $H_t<H_0$ indicates
enhancement of~$u_t$.

\begin{figure}
\includegraphics{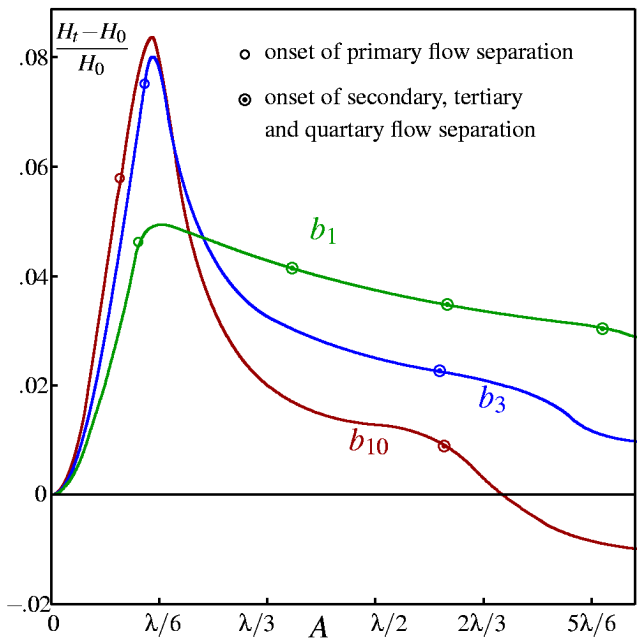}
\caption{\label{flowrate}Relative film elevation vs. amplitude of flows
 over different shapes.}
\end{figure}

In FIG.~\ref{flowrate} the relative film elevation~$(H_t-H_0)/H_0$
is plotted versus the amplitude~$A$ for the three different bottom
contours~$b_1$, $b_3$ and $b_{10}$.
This parameter study has been carried out with a fixed flow rate of
$\dot V=9\varrho g\lambda^3\sin\alpha/(8\eta)$, corresponding to
a reference thickness of $H_0=3\lambda/2$.
%
Additionally, the onset of primary
and further flow separation is indicated in the diagram.
\color{black}%
From the beginning up to an amplitude of about $\lambda/6$,
the film elevation is monotonously increasing for all three
different shapes, which indicates a reduction of the
mean transport velocity due to the bottom corrugations.
Note, that within this parameter regime no positive effect
can be expected since no vortices are present.
However, the curves reach maxima slightly after the
primary vortex generation and pass then into a
monotonous decrease due to reduction of friction by
vortices, which act like fluid roller bearings.
Obviously, both the height of the maximum as well as the decrease
after the vortex generation are more pronounced
for bottom shapes with sharper peaks.
For the curve associated to the bottom~$b_{10}$ the
film elevation becomes negative for high amplitudes, i.e.
the mean transport velocity exceeds the corresponding
mean transport velocity of the flow over a plane bottom.
At the highest amplitude considered in our calculations,
$A=0.9\lambda$, this enhancement of the mean transport
velocity reaches $0.88\%$.

The increase of the mean transport velocity indicates
an improved mass transport in the volume.
In contrast to this, the surface velocity remains nearly
unchanged.

\section{Conclusion}
%
 
%
The parameter studies on the three different shapes $b_1$,
$b_3$ and $b_{10}$ revealed a noticeable effect of the
bottom topography on the material transport in a creeping
film flow: For the shapes $b_1$ and $b_3$ we found a
decrease of the mean transport velocity compared to
the flow over a plane bottom, whereas for the shape~$b_{10}$
with the sharpest peaks an increase of the mean transport velocity
becomes apparent at sufficiently high amplitude.
Thus, a comparison to the 'shark skin effect',
which has been successfully applied to ships and airplanes for
drag reduction, is near at hand. The present effect, however, is
essentially different from the popular shark skin effect:
The increase of the mean transport velocity has been calculated
for creeping flows, whereas the shark skin effect occurs in
turbulent flow at Reynolds numbers $\approx 10^4$--$10^6$.
Furthermore, the rippled structures of the shark skin are
directed longitudinal to the flow, not transversal as it is
the case here. Finally, the responsible mechanism for the shark skin
effect is according to \cite{Nachtigall} and \cite{Dinkelacker02}
the control of the streamwise vortices in the turbulent flow,
which has been created by inertia.
In contrast to this, the rippled bottom structure in creeping
film flows enforce the creation of vortices which act on the
flow like a kind of 'fluid roller bearing'.
Nevertheless, a common feature of the shark skin effect
and the effect observed in the present paper is the reduction
of resistance in the flow by means of rippled wall structures. 

The highest relative increase of the mean transport velocity
is in our calculations $0.88\%$.
This is less than the
shark skin effect in turbulent flow, which is supposed
to be at most 10\%~\cite{Dinkelacker02}.
It is an open question up to which extend the value of 0.88\%
\color{black}
can be improved by varying the relevant parameters,
especially the bottom shape.



\bibliography{letter}

\end{document}